# The Early History of Microquasar Research

## I. Félix Mirabel

IAFE/CONICET/UBA, cc 67, suc 28, 1428, Buenos Aires Argentina
CEA/IRFU/SAp/Saclay, 91191 Gif-sur-Yvette Cedex, France

**Summary.** Microquasars are compact objects (stellar-mass black holes and neutron stars) that mimic, on a smaller scale, many of the phenomena seen in quasars. Their discovery provided new insights into the physics of relativistic jets observed elsewhere in the universe, and the accretion--jet coupling. Microquasars are opening new horizons for the understanding of ultraluminous X-ray sources observed in external galaxies, gamma-ray bursts of long duration, and the origin of stellar black holes and neutron stars. Microquasars are one of the best laboratories to probe General Relativity in the limit of the strongest gravitational fields, and as such, have become an area of topical interest for both high energy physics and astrophysics.
**Keywords:** Microquasar; Black Hole; General Relativity; Gravitation

## 1. Introduction

Microquasars are binary stellar systems where the remnant of a star that has collapsed to form a dark and compact object (such as a neutron star or a black hole) is gravitationally linked to a star that still produces light, and around which it makes a closed orbital movement. In this cosmic dance of a dead star with a living one, the first sucks matter from the second, producing radiation and very high energetic particles (Fig. 1). These binary star systems in our galaxy are known under the name of `microquasars' because they are miniature versions of the quasars (`quasi-stellar-radio-source'), that are the nuclei of distant galaxies harboring a super massive black hole, and are able to produce in a region as compact as the solar system, the luminosity of 100 galaxies like the Milky Way. Nowadays the study of microquasars is one of the main scientific motivations of the space observatories that probe the X-ray and gamma-ray Universe.

Despite the differences in the involved masses and in the time and length scales, the physical processes in microquasars are similar to those found in quasars. That is why the study of microquasars in our galaxy has enabled a better understanding of what happens in the distant quasars and AGN. Moreover, recent studies have revealed that microquasars may be related to two other high energetic objects: the non-identified gamma sources in our galaxy, and Gamma-Ray-Bursts, which are the most energetic phenomena in the Universe after the Big-Bang.

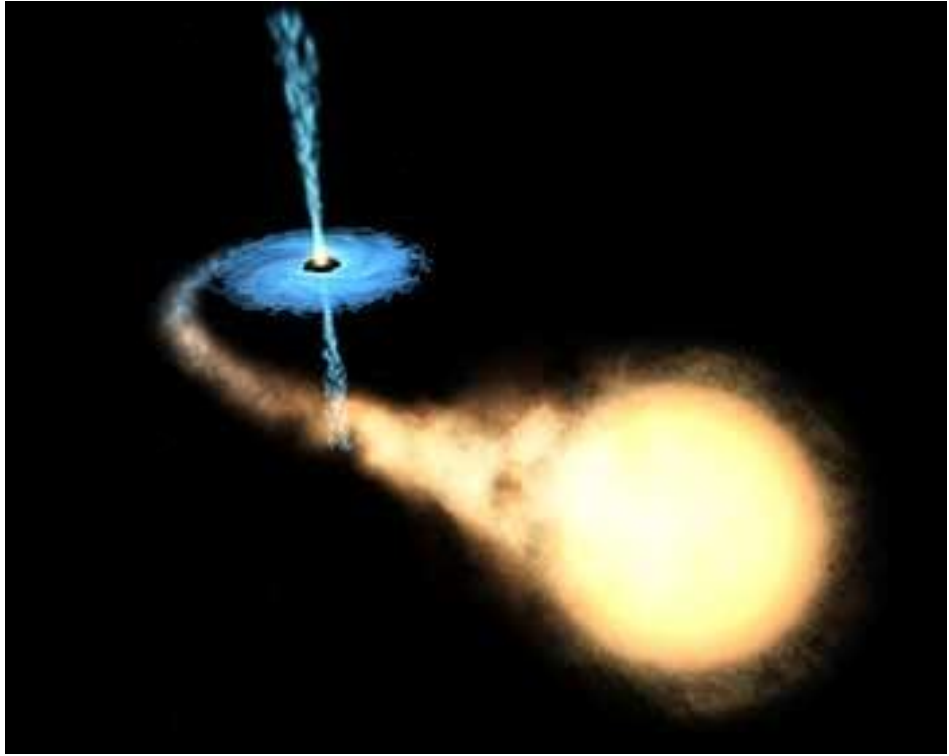

**Figure 1:** In our galaxy there exist binary stellar systems where an ordinary star gravitates around a black hole that sucks the outer layers of the star's atmosphere. When falling out to the dense star, the matter warms and emits huge amounts of energy as X-and gamma-rays. The accretion disk that emits this radiation also produces relativistic plasma jets all along the axis of rotation of the black hole. The physical mechanisms of accretion and ejection of matter are similar to those found in quasars, but in million times smaller scales. Those miniature versions of quasars are known under the name of `microquasars'. Image from joint press release by NASA, ESA and NRAO on the publication [20] of the runaway microquasar GRO J1655-40.

## 2. Discovery of microquasars

John Michell and Pierre-Simon Laplace first imagined compact and dark objects in the context of the classical concept of gravitation during the second half of the 18th century. In the 20th century, in the context of Einstein's General Relativity theory of gravitation, those compact and dark objects were named neutron stars and black holes. They were then identified in the sky, in the 1960s, as X-ray binaries. Indeed, those compact objects, when associated with other stars, are activated by the accretion of very hot gas that emits X- and gamma-rays. In 2002, Riccardo Giacconi received the Nobel Prize for the development of the X-ray Space Astronomy which led to the discovery of the first X-ray binaries [1]. In 1979, Margon et al. [2] found that a compact binary known as SS 433 was able to produce jets of matter. However, for a long time, people believed that SS 433 was a very rare object of the Milky Way and its relation with quasars was not clear since the jets of this object move only at 26% of the speed of light, whereas the jets of quasars can move at speeds close to the speed of light.

In the 1990s, after the launch of the Franco-Russian satellite GRANAT, growing evidences of the relation between relativistic jets and X-ray binaries began to appear. Its on-board telescope SIGMA was able to take X-ray and Gamma-ray images. It detected numerous black holes in the Milky Way. Moreover, thanks to the coded-mask-optics, it became possible for the first time to determine the position of gamma-ray sources with arcmin precision. This is not a very high precision for astronomers who are used to dealing with other observing techniques. However, in high-energy astrophysics it represented a gain of at least one order of magnitude. It consequently made possible the systematic identification of compact gamma ray sources in other wavelengths (radio, infrared and visible).

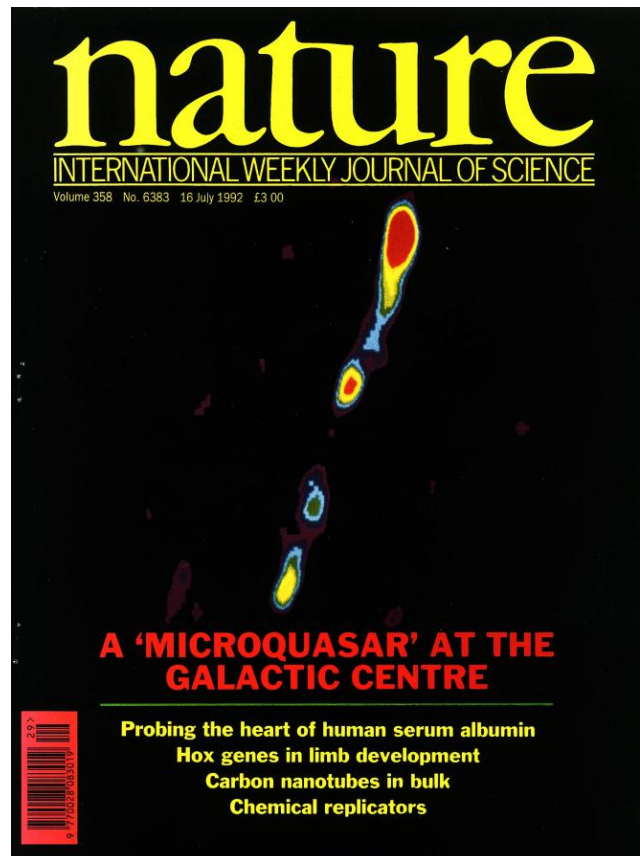

**Figure 2:** The British journal *Nature* announced on the 16th of July, 1992 the discovery of a microquasar in the galactic centre region [3]. The image on the cover is the synchrotron emission in radio wavelengths produced by relativistic particles jets ejected from some tens of kilometers to light-years of distance from the black hole.

GRANAT was launched in the same week that I left the California Institute of Technology to start working at the French Atomic Energy Commission on research projects on luminous infrared galaxies with the Infrared Space Observatory of the European Space Agency. My colleagues Jacques Paul and Bertrand Cordier talked to me about X-ray and Gamma-ray sources that with SIGMA/GRANAT one was able to localize with an unprecedented precision. In order to determine the nature of those X-ray binaries, a higher precision was, however, necessary (a few tens of arc-second). However, sources that produce highly energetic photons should also produce highly energetic particles that should then produce synchrotron radiation when accelerated by a magnetic field. This is why I proposed to Luis Felipe Rodriguez,---one of the best specialists in interferometry at radio wavelengths, who was at that time director of the Institute of Astronomy of the Universidad Autonoma de Mexico---to perform a systematic search of synchrotron emissions in X-ray binaries with the Very Large Array (VLA) of the National Radio Astronomy Observatory of the USA.

In 1992, using quasi-simultaneous observations from space with GRANAT and from the ground with the VLA, we determined with a precision of sub-arc-seconds the position of the radio counterpart of an X-ray source named 1E1740.7-2942, which had been detected with the satellite Einstein. With GRANAT this object was identified as the most luminous persistent source of soft gamma-rays in the Galactic center region. Moreover, its luminosity, variability and the X-ray spectrum were consistent with those of an accretion disk gravitating around a stellar mass black hole, like in Cygnus X-1. The most surprising finding with the VLA was the existence of well collimated two-sided jets that seem to arise from the compact radio counterpart of the X-ray source [3]. These jets of magnetized plasma had the same morphology as the jets observed in quasars and radio galaxies. When we published those results, we employed for the first time the term microquasar to define this new X-ray source with relativistic jets in our Galaxy. This term appeared on the front page of the British journal *Nature* (see Fig. 2), which provoked multiple debates. Today the concept of microquasar is universally accepted and used in scientific publications. Poetry in science may be the quest for words that stimulate the imagination and open new research perspectives.

Before the discovery of its radio counterpart, 1E1740.7-2942 had been proposed as the compact source of a possible variable component in the 511-keV electron-positron annihilation radiation observed since the 1970s from the Galactic Centre region of our Galaxy. An editorial article in the journal Physics Today nicknamed it as the "Great Annihilator". Although accreting stellar black holes could be an important source of positrons that annihilate with electrons in the interstellar medium, no variability and spatial asymmetry in the 511 keV emission from the Galactic Centre region has been confirmed by more recent observations with the satellite Integral.

## 3. Discovery of super-luminal motions

Until now about 40 microquasars have been discovered in our galaxy [4]. If the analogy between microquasars and quasars is correct, it must be possible to observe super-luminal apparent motions in Galactic sources. However, super-luminal apparent motions had been observed only in the neighborhood of super-massive black holes, namely, in quasars. The illusion of super luminous motions is due to the relativistic aberration of the radiation produced by particles that move at speeds close to the speed of light. In 1E 1740.7-2942 we could not discern motions, as in that persistent source of gamma rays the flow of particles is semi-continuous. The only possibility of knowing if super-luminal apparent movements exist in microquasars was through the observation of a discreet and very intense ejection in an X-ray binary. This would allow us to follow the displacement in the firmament of discrete plasma clouds. Indeed in 1992, Castro-Tirado [5], from the Andalusian Institute of Astrophysics and collaborators, discovered with the GRANAT satellite, a new source of X-rays with such characteristics, denominated GRS 1915+105. Then with Rodriguez we began with the VLA a systematic observing campaign of that new object in the radio domain. In collaboration with the young French Pierre-Alain Duc (CNRS-France) and Sylvain Chaty (Paris University) we performed the follow-up of this source in the infrared with telescopes of the Southern European Observatory, and at 4200 meters in Mauna Kea, Hawai.

Since the beginning, GRS 1915+105 exhibited unusual properties. The observations in the optical and the infrared showed that this X-ray binary was very absorbed by the interstellar dust along the line of sight in the Milky Way, and that the infrared counterpart was varying rapidly as a function of time. Moreover, the radio counterpart seemed to change its position in the sky, so that at the beginning we did not know if those changes were due to radiation reflection or refraction in an inhomogeneous circumstellar medium ("Christmas tree effect"), or rather due to the movement at very high speeds of jets of matter. For two years we kept on watching this X-ray binary without exactly understanding its behavior. However, in March 1994, GRS 1915+105 produced a violent eruption of X- and gamma-rays, followed by a bipolar ejection of unusually bright plasma clouds, whose displacement in the sky could be followed during two months. From the amount of atomic hydrogen impressed in the strong radiation we could infer that the X-ray binary stands at about 40000 light years from the Earth. This enabled us to know that the movement in the sky of the ejected clouds implies apparent speeds greater than the speed of light.

The discovery of these super-luminal apparent movements in the Milky Way was announced in *Nature* [6] (Fig. 3). This constituted a full confirmation of the hypothesis, that we had proposed two years before, on the analogy between microquasars and quasars. With Rodriguez we formulated and solved the system of equations that describe the observed phenomenon. The apparent asymmetries in the brightness and the displacement of the two plasma clouds could naturally be explained in terms of the relativistic aberration in the radiation of twin plasma clouds ejected in an antisymmetric way at 98% of the speed of light [6]. The solving of those equations was the first problem posed to the most brilliant students of the world in the International Physic Olympiads in 2000. The super-luminal motions we observed in the period of 1994 to 1998 with the VLA [7] were years later, confirmed by observations of Fender et al. [8] using MERLIN.

Using the Very Large Telescope of the European Southern Observatory, Greiner and collaborators [9] were able to determine the orbital parameters of GRS 1915+105, concluding that it is a binary system constituted by a black hole of 14 solar masses accompanied by a star of 1 solar mass. The latter has become a red giant from which the black hole sucks matter under the form of an accretion disk (see Fig. 1).

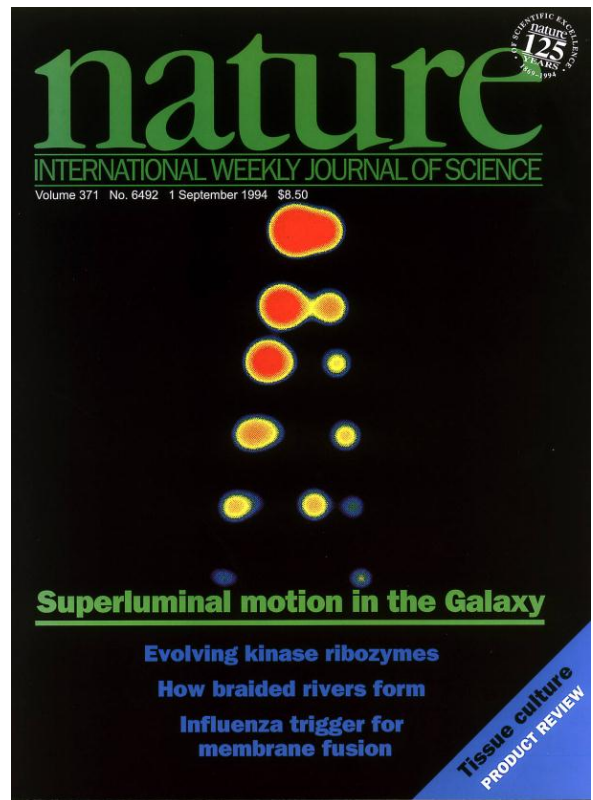

**Figure 3.** The journal *Nature* announces on the 1st of September, 1994 the discovery of the first Galactic source of super-luminal apparent motions [6]. The sequence of images on the cover shows the temporal evolution in radio waves of a pair of plasma clouds ejected from black hole surroundings at a speed of 98% the speed of light.

## 4. Observation in real time of the accretion disk-jet coupling

The association of bipolar jets and accretion disks seems to be a universal phenomenon in quasars and microquasars. The predominant idea is that matter jets are driven by the enormous rotation energy of the compact objects and accretion disks that surround it. Through magneto-hydrodynamic mechanisms, the rotation energy is evacuated through the poles by means of jets, as the rest can fall towards the gravitational attraction centre. In spite of the apparent universality of this relationship between accretion disks and bipolar, highly collimated, jets, it had never been confirmed by direct observations in real time.

Since the scales of time of the phenomena around black holes are proportional to their mass [4], the accretion-ejection coupling in stellar-mass black holes can be observed in intervals of time that are millions of time smaller than in AGN and quasars. Because of the proximity, the frequency and the rapid variability of energetic eruptions, GRS 1915+105 became the most adequate object to study the connection between instabilities in the accretion disks and the genesis of bipolar jets.

Finally in 1997, after several attempts, we could observe [10] on an interval of time shorter than an hour, a sudden fall in the luminosity in X and soft gamma-rays, followed by the ejection of jets, first observed in the infrared, then at radio frequencies (see Fig. 4). The abrupt fall in X-ray luminosity could be interpreted as the silent disappearance of the warmer inner part of the accretion disk beyond the horizon of the black hole. A few minutes later, fresh matter coming from the companion star comes to feed again the accretion disk, which must evacuate part of its kinetic energy under the form of bipolar jets. When moving away, the plasma clouds dilute, becoming transparent to its own radiation, first in the infrared and then in radio frequency. The observed interval of time between the infrared and radio peaks is consistent with the model by van der Laan [11] for the time delay of outbursts as a function of wavelength originally proposed for extragalactic radio sources.

Based on the observations of GRS 1915+105 and other X-ray binaries, Fender, Belloni and Gallo [12] proposed a unified semiquantitative model for disk-jet coupling in black hole X-ray binary systems that relate different X-ray states with radio states, including the compact, steady jets associated with low-hard X-ray states, that had been imaged by Dhawan, Mirabel and Rodriguez [13] using the Very Long Baseline Array of the National Radio Astronomy Observatory.

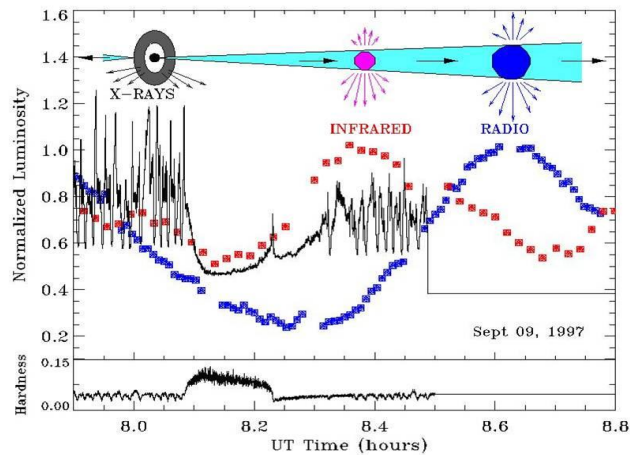

**Figure 4**. Accretion disk - jet coupling observed for first time in real time in a microquasar, simultaneously in the X-rays, the infrared and radio wavelengths [10]. The ejection of relativistic jets takes place after the evacuation and /or dissipation of matter and energy, at the time of the reconstruction of the inner side of the accretion disk, corona or base of the jet.

A similar process was expected to be observed in quasars, but on time scales of years, because in the context of the analogy between quasars and microquasars (4), the time scale of physical processes in the surroundings of black holes is proportional to their masses. In fact, after three years of multi-wavelength monitoring, an analogous sequence of X-ray emission dips followed by the ejection of bright super-luminal knots in radio jets was reported by Marscher et al. [14] in the active galactic nucleus of the galaxy 3C120. The mean time between X-ray dips was of the order of years, which as expected appears to scale roughly with the mass of the black hole.

## 5. Can we prove the existence of black holes?

The horizon is the basic concept that defines a black hole: a massive object that consequently produces a gravitational attraction in the surrounding environment, but that has no material border. In fact, an invisible border in the space-time, which is predicted by general relativity, surrounds it. This way, matter could go through this border without being rejected, and without losing a fraction of its kinetic energy in a thermonuclear explosion, as would happen if the compact object were a neutron star instead of a black hole. In fact, as shown in Figure 4, the interval of time between the sudden drop of the flux and the spike in the X-ray light curve that marks the onset of the jet signaled by the initial rise in the infrared synchrotron emission, is of a few minutes, an order of magnitude larger than the dynamical time of the plasma in the inner accretion disk. In other words, it is as if the matter that falls into the black hole disappeared mysteriously and the hot gas that was producing the X-ray emission would leave our observable Universe.

So, have we proved with such observations the existence of black holes? Indeed, we do not find any evidence of material borders around the compact object that creates gravitational attraction. However, the fact that we do not find any evidence for the existence of a material surface does not imply that it does not exist. That means that it is not possible to prove the existence of black holes using the horizon definition. According to Saint Paul, ``*faith is the substance of hope for: the evidence of the not seen*''. That is why for some physicists, black holes are just objects of faith. Perhaps the intellectual attraction of these objects comes from the desire to discover the limits of our Universe. In this context, studying the physical phenomena near the horizon of a black hole is a way of approaching the ultimate frontiers of our observable Universe.

For an external observer, however, black holes are the simplest objects in physics since they are fully described with only 3 parameters: mass, rotation and charge. The masses of black holes gravitating in binary systems can be estimated with Newtonian physics. However, the rotation is much more difficult to estimate despite it being probably the main driver in the production of relativistic jets. The discovery of microquasars opened the possibility of measuring the rotation of black holes using quasi-periodic oscillations with a constant maximum frequency that is observed in the X-rays. The side of the accretion disk that is closer to the black hole is hotter and produces huge amounts of thermal X and gamma radiations and is also affected by the strange configuration of space-time. Indeed, next to the black hole, space-time is curved by the black hole mass and dragged by its rotation. This produces vibrations that modulate the X-ray emission. The study of those X-ray vibrations has enabled us to discover that the microquasars that produce the most powerful jets are indeed those that are rotating fastest. These pseudo-periodic oscillations in microquasars are, moreover, one of the best methods today to probe by means of observations, General Relativity theory in the limit of the strongest gravitational fields.

There have been reports of analogous oscillations in the infrared range, coming from the super massive black hole in the centre of the Milky Way. The quasi-periods of the oscillations (a few milliseconds for the microquasars X-ray emission and a few tens of minutes for the Galactic centre black hole infrared emission) are proportionally related to the masses of the objects, as expected from the physical analogy between quasars and microquasars.

Comparing the phenomenology observed in microquasars to that in black holes of all mass scales, several correlations among observables such as among the radiated fluxes, quasi-periodic oscillations, flickering frequencies, etc., are being found and used to derive the mass and spin, which are the fundamental parameters that describe astrophysical black holes.

## 6. Extragalactic microquasars

Have microquasars been observed beyond the Milky Way galaxy? X-ray satellites are detecting far away from the centers of external galaxies, large numbers of compact sources called `ultraluminous X-ray sources', because their luminosities seem to be larger than the Eddington limit for a stellar-mass black hole. Although a few of these sources could be black holes of intermediate masses of a few hundreds to thousands solar masses, it is believed that the large majority are stellar-mass black hole binaries where the X-ray radiation is anisotropic. In fact, the Galactic microquasar SS 433 has near super-Eddington kinetic luminosity and if it were observed from another point of view in the Galaxy it would very likely appear as an ultraluminous X-ray source.

Since the discovery of quasars in 1963, it was known that some quasars could be extremely bright and produce high energetic emissions with rapid variability in short times. These particular quasars are called blazars and it is thought that they are simply quasars whose jets are exactly in the Earth's direction. The Doppler effect produces thus an amplification of the signal and a shift into higher frequencies. With Rodriguez we imagined in 1999 the existence of microblazars, that is to say X-ray binaries where the emission is also in the Earth's direction [7]. Microblazars may have been already observed but the fast variations caused by the contraction of the time scale in the relativistic jets, make their study very difficult.

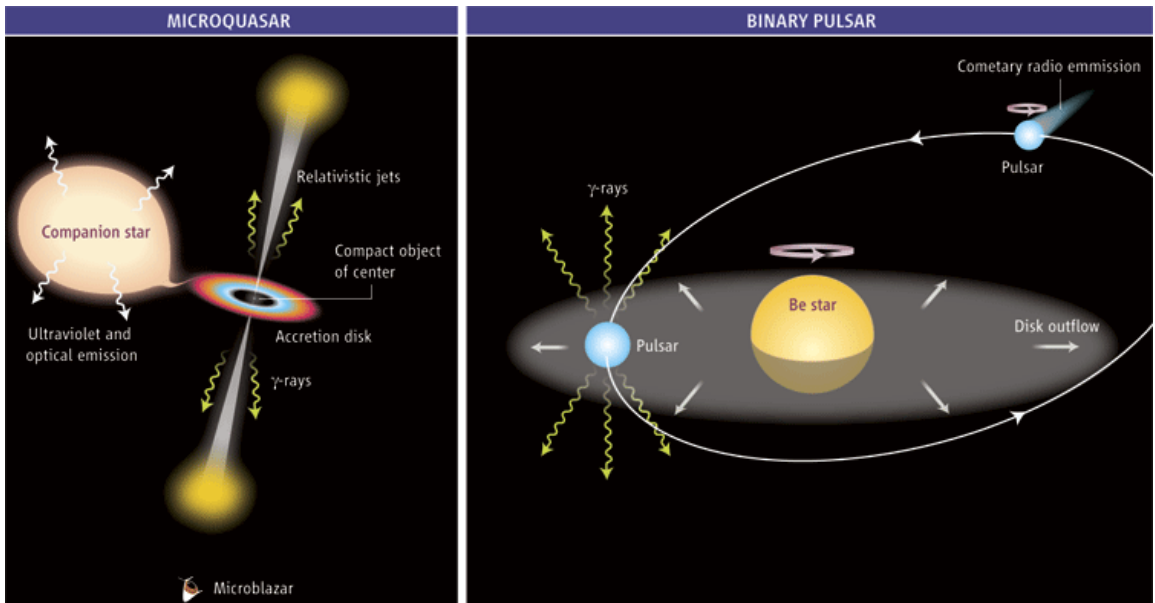

**Figure 5.** Alternative models for very energetic gamma-ray binaries [15,16]. Left: microquasars are powered by compact objects (neutron stars or stellar-mass black holes) via mass accretion from a companion star. The interaction of collimated jets with the massive outflow from the donor star can produce very energetic gamma-rays by different alternative physical mechanisms depending on whether the jets are baryonic or purely leptonic. Right: pulsar winds are powered by rotation of neutron stars; the wind flows away to large distances in a comet-shape tail. Interaction of this wind with the companion-star outflow may produce very energetic gamma-rays.

## 7. Very energetic gamma-ray emission from compact binaries

Very energetic gamma-rays with energies greater than 100 gigaelectron volts have recently been detected with ground based telescopes from a few high mass compact binaries [15]. These have been interpreted in the context of the two models represented in Fig. 5. In two of these sources the gamma radiation is correlated with the orbital phase of the binary, and therefore consistent with the idea that the very high energy radiation may be produced by the interaction of a pulsar wind with the mass outflow from the massive companion star. It is a current question whether the gamma-ray binaries being observed by the satellites Fermi and Agile are microquasars where the radiation in the gamma domain is produced by the interaction of the jets with the outflow from the massive donor star, as proposed by the teams led by Gustavo Romero et al. from the CONICT in Argentina and Josep Maria Paredes from the University of Barcelona. In a recent publication [16] I have proposed the perspectives in this area of research opened by new theoretical and observational results.

## 8. Formation of stellar black holes

Runaway microquasars have been studied using multiwavelength data obtained with a diversity of observational techniques. The team from the University of Barcelona found that the micro quasar LS 5039 has been ejected out of the Milky Way at high speed, therefore the compact object formation must have been caused by the explosion of a very energetic supernova. Studies of microquasars kinematics performed today are the beginning of what could be called `Galactic Archaeology'. Like archaeologists, studying fossils of progenitors of the present stars in our galaxy, astrophysicians can know the birth, life and death of those past stars generations. With this spirit, in collaboration with Irapuan Rodrigues from South Rio Grande University in Brazil, we found runaway black holes moving at high speed, some moving like globular clusters in the Galactic halo [17] (Fig. 6). It remains an open question whether this halo black hole was kicked out from the Galactic plane by a natal explosion, or is the fossil of a star that formed in a globular cluster more than 7 billions of years ago, before the spiral disk of stars, gas and dust of the Milky Way was formed.

It is of interest to know why some massive stars finish their life as neutron stars whereas others form black holes. It is also interesting to know if the massive stars that form black holes always die in a violent way, exploding like supernovae. To answer to these questions, microquasars kinematics is a complementary approach to the study of Gamma-Ray-Bursts. When a binary system of massive stars is still gravitationally linked after the supernova explosion of one of its components, the mass centre of the system acquires an impulse, whatever matter ejection it is, symmetric or asymmetric. Then according to the microquasar movement we can investigate the origin and the formation mechanism of the compact object. Our preliminary results on the kinematics of the X-ray binaries suggest that low mass black holes are formed by a delayed collapse of a neutron star, whereas stellar black holes with a mass equal or greater than 10 solar masses are probably the result of the direct collapse of massive stars. In fact, we have found that Cygnus X-1, the first black hole discovered, was formed in a dark way [18].

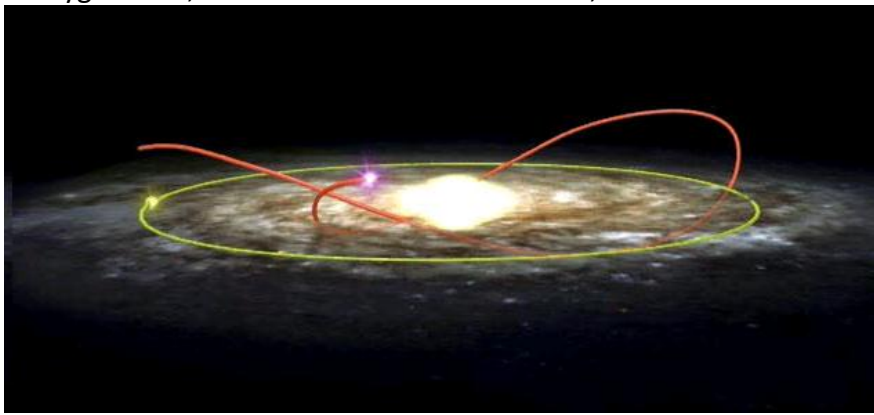

**Figure 6:** A wandering black hole in the Galactic halo [17]. The trajectory of the black hole for the last 230 million years is represented in red. The yellow dot represents the Sun.

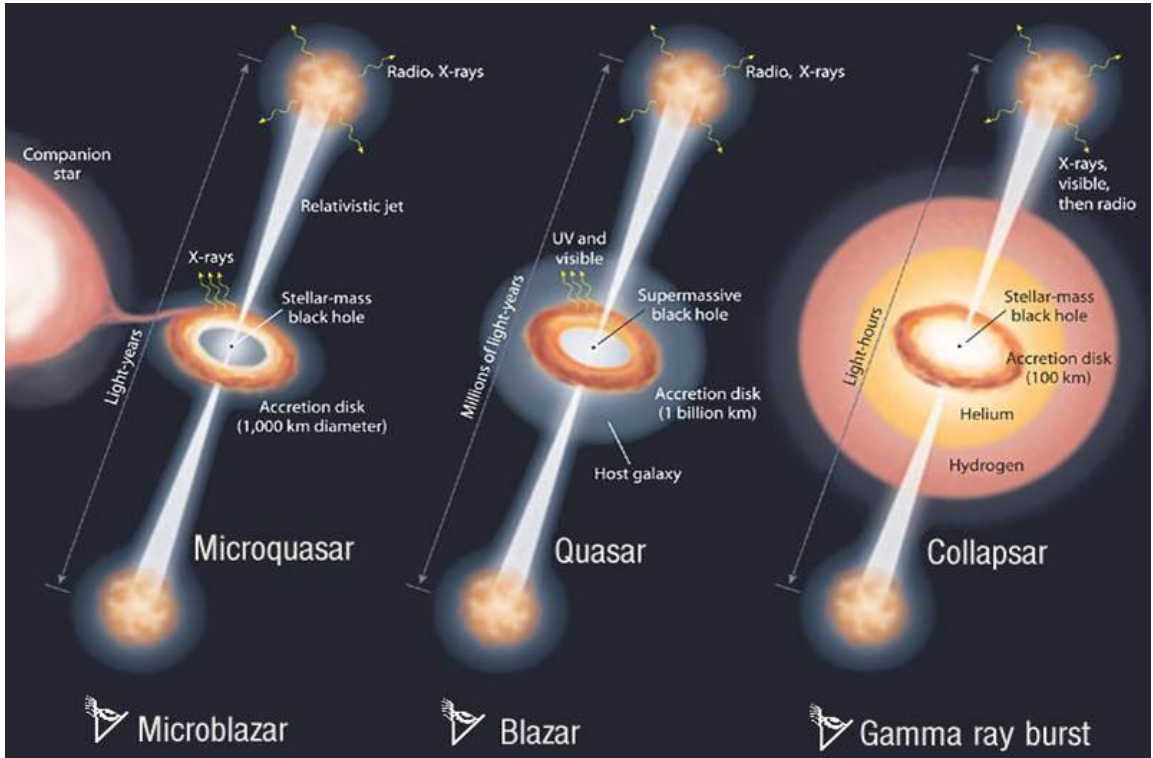

**Figure 7:** The same physical mechanism can be responsible for three different types of objects: microquasars, quasars and massive stars that collapse ('collapsars') to form a black hole producing Gamma-Ray-Bursts [19]. Each one of these objects contains a black hole, an accretion disk and relativistic particles jets. Quasars and microquasars can eject matter several times, whereas the collapsars form jets only once. When the jets are aligned with the line of sight of the observer these objects appear as microblazars, blazars and gamma bursts, respectively.

## 9. Long gamma-ray bursts

The sources of Gamma-Ray-Bursts (GRBs) that last from a few seconds to minutes are the most energetic explosions that can be observed in the whole Universe. With the young French astronomer Emeric Le Floc'h we found that these GRBs are preferentially produced in host galaxies of low metallicity. GRBs are believed to be produced by ultra relativistic jets generated in a massive star nucleus when it catastrophically collapses to form a black hole. Gamma-Ray-Bursts are highly collimated jets and with Luis Felipe Rodriguez [19] we have proposed that there may be a unique universal mechanism to produce relativistic jets in the Universe, suggesting that the analogy between microquasars and quasars can be extended to the Gamma-Ray-Bursts sources, as illustrated in the diagram of Fig. 7.

## 10. Conclusions

Black hole astrophysics is presently in an analogous situation to that of stellar astrophysics in the first decades of the 20th century. At that time, well before reaching the physical understanding of the interior of stars and the way by which they produce and radiate their energy, empirical correlations such as the HR diagram were found and used to derive fundamental properties of the stars, such as the mass. Similarly, in black hole astrophysics, empirical correlations among observables of black holes of all mass scales are being used to derive their mass and spin, which are the fundamental parameters that describe astrophysical black holes.

In summary, this research area on microquasars that we initiated 20 years ago has become one of the most important areas in high energy astrophysics. Since their discovery there have been eight international workshops on microquasars: 4 in Europe, 2 in America and 2 in Asia. They are being attended by 100-200 young scientists who, with their work on microquasars, are contributing to open new horizons in the common ground of high energy physics and modern astronomy.